\documentclass{article}

\usepackage[preprint]{neurips_2023}

\usepackage[utf8]{inputenc} 
\usepackage[T1]{fontenc}    
\usepackage{hyperref}       
\usepackage{url}            
\usepackage{booktabs}       
\usepackage{amsfonts}       
\usepackage{nicefrac}       
\usepackage{microtype}      
\usepackage{xcolor}         
\usepackage{subcaption}
\usepackage{graphicx}

\title{A Comparative Study of Machine Learning Models Predicting Energetics of Interacting Defects}

\author{%
  Hao Yu\\ 
  Department of Electrical and Computer Engineering\\
  Boston University\\
  Boston, MA 02215 \\
  \texttt{imhaoyu@bu.edu} \\
}

\begin{document}

\maketitle

\begin{abstract}
Interacting defect systems are ubiquitous in materials under realistic scenarios, yet gaining an atomic-level understanding of these systems from a computational perspective is challenging -  it often demands substantial resources due to the necessity of employing supercell calculations. While machine learning techniques have shown potential in accelerating materials simulations, their application to systems involving interacting defects remains relatively rare. In this work, we present a comparative study of three different methods to predict the free energy change of systems with interacting defects. We leveraging a limited dataset from Density Functional Theory(DFT) calculations to assess the performance models using materials descriptors, graph neural networks and cluster expansion. Our findings indicate that the cluster expansion model can achieve precise energetics predictions even with this limited dataset. Furthermore, with synthetic data generate from cluster expansion model at near-DFT levels, we obtained enlarged dataset to assess the data requirements for training accurate prediction models using graph neural networks for systems featuring interacting defects. A brief discussion of the computational cost for each method is provided at the end. This research provide a preliminary evaluation of applying machine learning techniques in imperfect surface systems.
\end{abstract}

\section{Introduction}
\begin{figure}[h]
    \centering
    \includegraphics[width=0.5\linewidth]{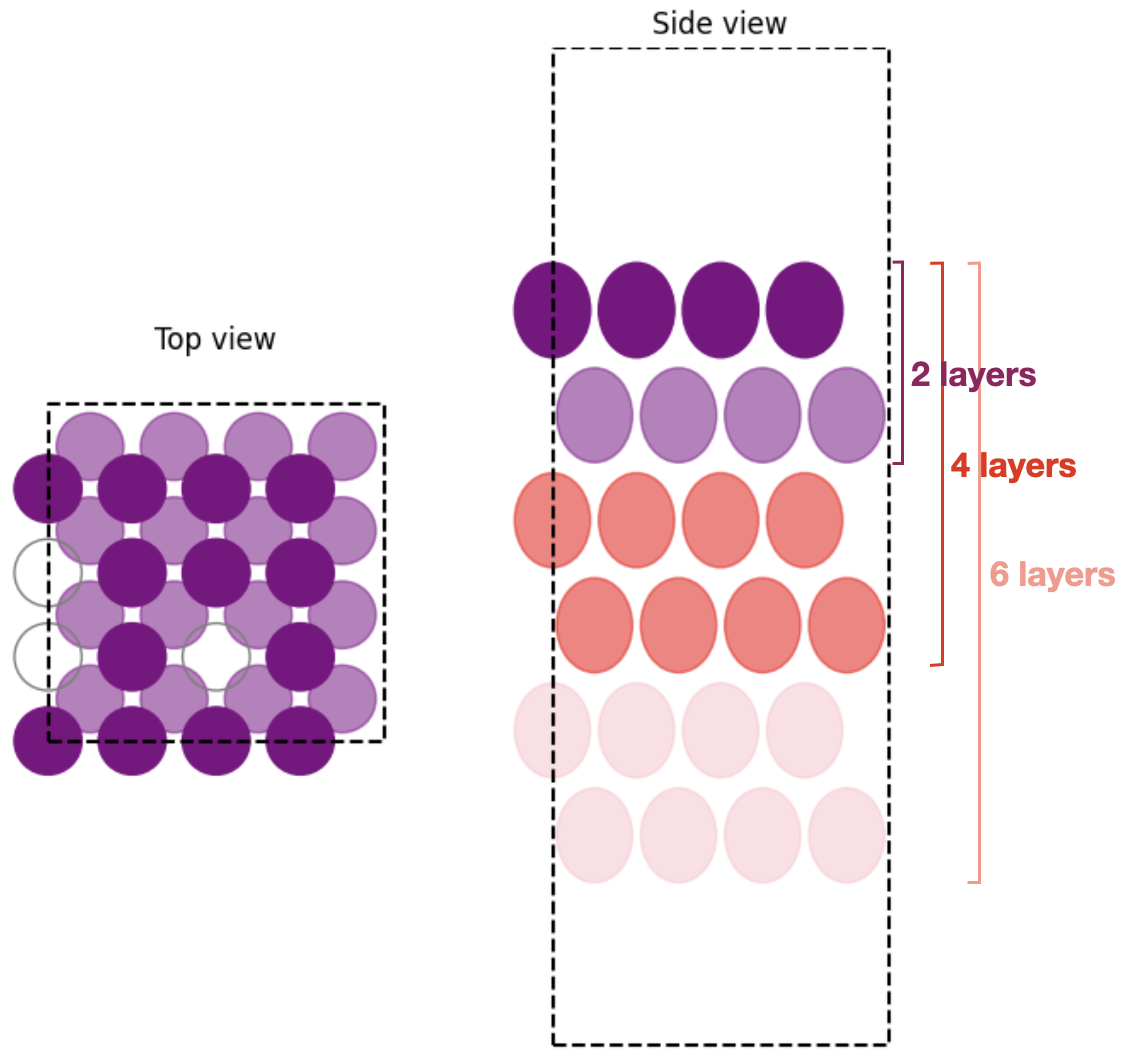}
    \caption{Example of lithium slab with defects on surface. Left: Top view is looking at the surface with defects (represented by gray circles). Circles filled with purple represent the surface lithium atoms, while circles with half filled purple represent the lithium atoms in the second layer. Dashed square indicate the unit cell. Right: Side view of a lithium slab. Surface atoms represented with filled purple circles. Atoms at first 2 layers match the ones on the left figure. Dashed square is unit cell box. We tried different representations by keeping atoms with 2/4/6 layers in the structure.}
    \
    \label{fig:lithium}
\end{figure}

Defects in materials profoundly influence their properties. Understanding defects interactions is essential for optimizing material properties. Phenomena arising from defect interactions can occur from atomic-level to mesoscale and modeling across the scales is computationally challenging. Recently, machine learning potentials have shown success in accelerating materials simulations for molecules and crystals\citep{doi:10.1146/annurev-physchem-042018-052331}. However, surface systems with a number of defects are rarely studied. Except defects interactions are complex, simulations of surface systems are costly since calculation for large supercells are needed \citep{SUN201353}. These prevent the progress of studies on complex phenomena in surface systems. \\
To explore the potential of applying machine learning techniques to study interacting defects on surfaces at atomic level, we provided as case study of single-type defects on pure element crystal. Driven from possible applications in real-life problem related to lithium battery\citep{Liu_2017}, we focus on different numbers of vacancies on (100) surface of lithium (Figure \ref{fig:lithium}). We defined the defect concentration on surface as 
\begin{equation}
    \textrm{defect concentration}=\frac{\textrm{number of vacancies}}{\textrm{sites of perfect crystal on surface}} \times 100\%,
\end{equation}
The prediction task is, given the structure of defects distribute on surface to predict the free energy change $\Delta G$ when vacancies present on the surface. This free energy change is defined as 
\begin{equation}
    \Delta G= E_{\textrm{vac}}+nE_{\textrm{Li}}-E_{\textrm{Li slab}},
\end{equation}
where $E_{\textrm{vac}}$ is the ground state energy of lithium surface with defects from Density Functional Theory calculation(DFT), $n$ is the number of vacancies in this system and $E_{\textrm{Li}}$ is the chemical potential of lithium (atomic energy of lithium in its most stable form), and $E_{\textrm{Li slab}}$ is the ground state energy of a pristine lithium supercell. \\
Our initial data set calculated by DFT has 88 structures with concentration from 6.25\% to 50\%. We split the training set and test set by concentration to make the training set and test set dissimilar both in configurations and concentrations. The training set contains structures with concentration less than 40\%, and there are 73 structures. The test set contains the rest 15 structures with concentration higher than 40\%. We first study the model performance on initial test data with three possible approaches. Cluster expansion(CE) is the the approach we showed it can predict the target at high accuracy. We further generate more surface systems with different defect configurations and their energy using this cluster expansion model trained with initial data, and use these configuration-energy pairs to test the potential of applying graph neural network(GNN) in this type of systems. 
 Contributions of this work include (1) Qualitative reveal the limitation of machine learning potentials for systems with interacting defects and provide a model to alleviate; (2) Establish a computationally feasible way to test the potential of GNN for surface with intereacting defects.

\section{Methodology}
\subsection{Atomic Representations}
Within the machine learning framework of predicting property $y\in \mathbb{R}$ of a material system $(\mathbf{Z},\mathbf{r})$, where $\mathbf{Z}\in \mathbb{R}^{N\times 3}$ is the matrix of atomic positions and $\mathbf{r}\in \mathbb{N}^N$ is atomic numbers, it is assumed that the property $y$ has contribution from each atom in the system\citep{bader}. It can be formulated into regression task in two ways, (1) transform $(\mathbf{Z},\mathbf{r})$ to descriptors and use ML algorithms learn from these representations; (2) transform $(\mathbf{Z},\mathbf{r})$ as a graph $\mathcal{G}$ with $N$ nodes and edges represented in adjacency matrix $\mathbf{A}\in\mathbb{R}^{N\times N}$ and then apply graph neural network. Coulomb matrix (CM), Smooth Overlap of Atomic Positions(SOAP) and Many-body Tensor Representation (MBTR) are commonly used descriptors \citep{10.1063/5.0151031,HIMANEN2020106949}. Descriptors are encoded and limited within domain-knowledge from experts, it is effective when training data is little. Graphs are naturally good representation of atomic structures, and graph neural networks are used for tasks where graph-structured data has inherent relationships. To obtain a good GNN, large amount of data related to this task is needed. We used the framework provided by the open catalyst project\citep{ocp_dataset} conduct experiments. 

\subsection{Configurational Cluster Expansion}
 When the property of a material system is related to the configurational degrees of freedom, configurational cluster expansion is a way to build a model that can extrapolate predictions to unseen configurations. In our case, The lithium surface systems we consider only differs at surface configurations and there are $2^N$ configurations, $N$ is the sites of perfect crystal on surface. This collection is denoted as $\vec{\sigma}=(\sigma_1,...,\sigma_n,...,\sigma_N)$ The energy dependency with configurations in cluster expansion can be formulated with $E(\mathbf{\vec{\sigma}})=\sum_{\alpha}V_{\alpha}\Phi_{\alpha}(\vec{\sigma}),$
 where $\alpha$ refers to the point, pair, triplet, etc. clusters within the crystal, $V_\alpha$ is the effective interaction strength of a cluster and $\Phi_{\alpha}$ is the crystal basis function.
We can obtain these quantities when we construct a cluster expansion model provided few structures in the configurational space\citep{icet}. 

\section{Results}
\begin{figure}[h]
  \centering
  \begin{subfigure}{\textwidth}
      \includegraphics[width=\linewidth]{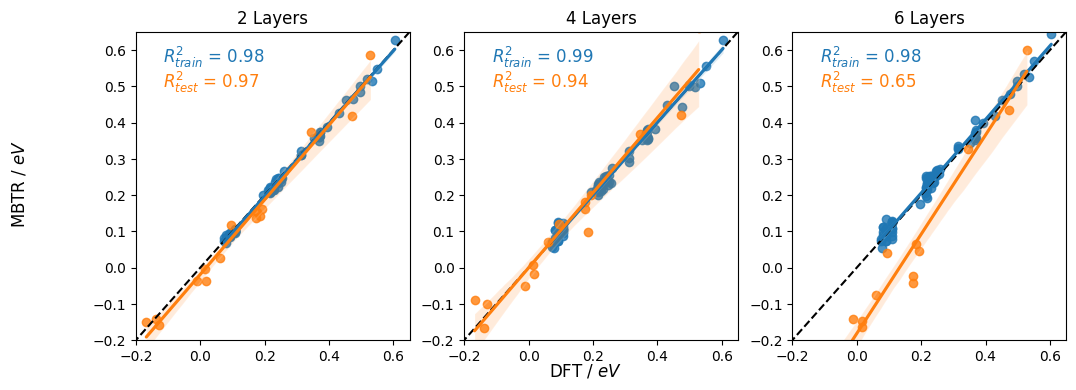}
  \end{subfigure}

  \begin{subfigure}{\textwidth}
      \includegraphics[width=\linewidth]{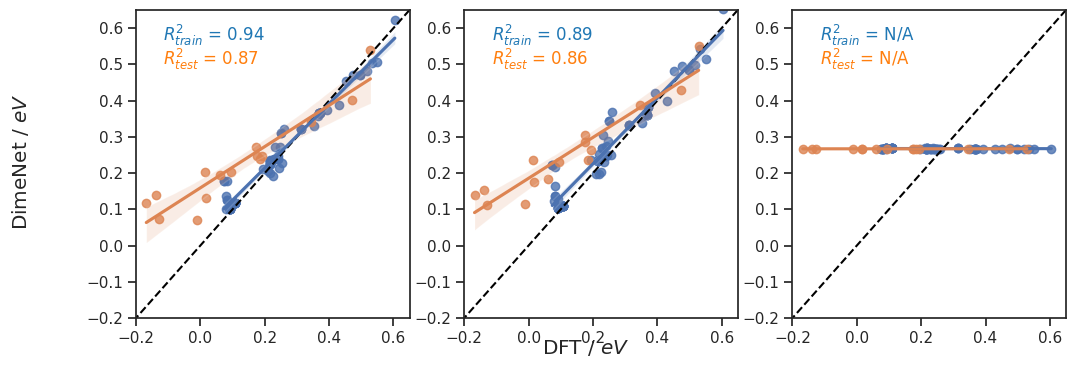}
  \end{subfigure}

  \begin{subfigure}{\textwidth}
      \includegraphics[width=\linewidth]{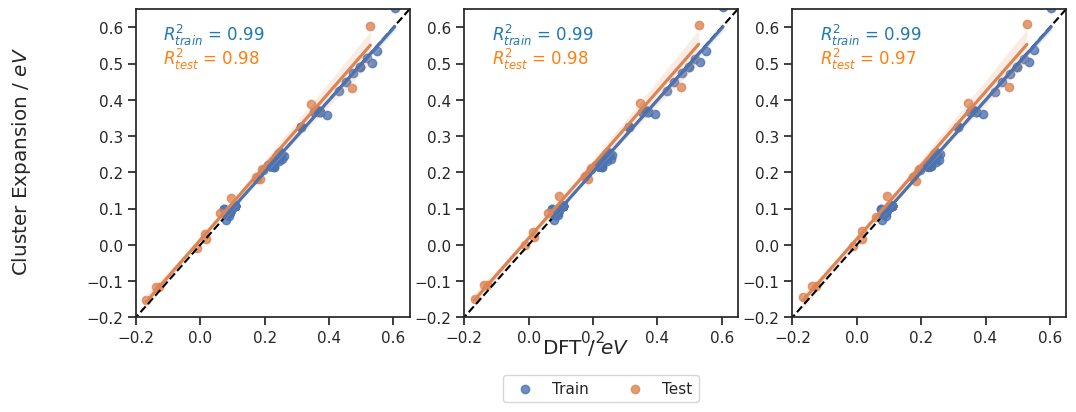}
  \end{subfigure}
  \caption{Train and Test predictions with initial small data set.}
  \label{fig1}
\end{figure}
\begin{table}[h]
  \caption{Statistics of Model Predictions}
  \label{table1}
  \centering
  \begin{tabular}{llll}
    \toprule
      & \multicolumn{3}{c}{MAE (Train/Test in $eV$)}  \\
    \cmidrule(r){2-4}
Model Name/ Layers  &2 &4 & 6\\
    \midrule
    MBTR  & 0.006/0.032  & 0.012/0.038&0.014/0.147\\
    DimeNet    & 0.020/0.117 & 0.028/0.138    & 0.117/0.219 \\
    Cluster Expansion    & 0.007/\textbf{0.021} &0.007/0.024  &0.007/0.024 \\
    \bottomrule
  \end{tabular}
\end{table}
The models were trained on systems with low defect concentrations and then tested on systems with high defect concentrations. We explored 3 kinds approaches: (1) descriptors + ML algorithms, (2) graph neural networks, (3) cluster expansion. For the first two approaches, we presented the prediction statistics of the most representative models, they are (1) MBTR\citep{Huo_2022} with linear regression, (2) DimeNet\cite{gasteiger2022directional}. We tested the best number of atoms to include in representation and the results in mean absolute error(MAE) and coefficient of determination($R^2$) are in Table \ref{table1} and Figure \ref{fig1}. All the models training with representations building from 2 layers of atoms, which are surface and subsurface atoms, have the lowest test MAE. Cluster expansion model can predict the energy with high concentration defects most accurately and achieving negligible error compare to DFT level accuracy. Linear regression model with MBTR also has low test MAE, however, the prediction error is higher when the concentration of defect is higher. DimeNet has the worst performance compare to the last two models, this is due to the training data is limited. Since the most important factor that impact the energy of this kind of extended system is surface configuration, when building a prediction model, information about the atoms far from surface makes less contribution in determining the free energy change of the system. This match well with the results in Figure \ref{table1} comparing column-wise. Notably, in graph neural network, if we represented the supercell structures as graphs while only surfaces differ, the model will treat every structure the same. 

\begin{figure}
    \centering
    \begin{subfigure}{\textwidth}
        \includegraphics[width=\linewidth]{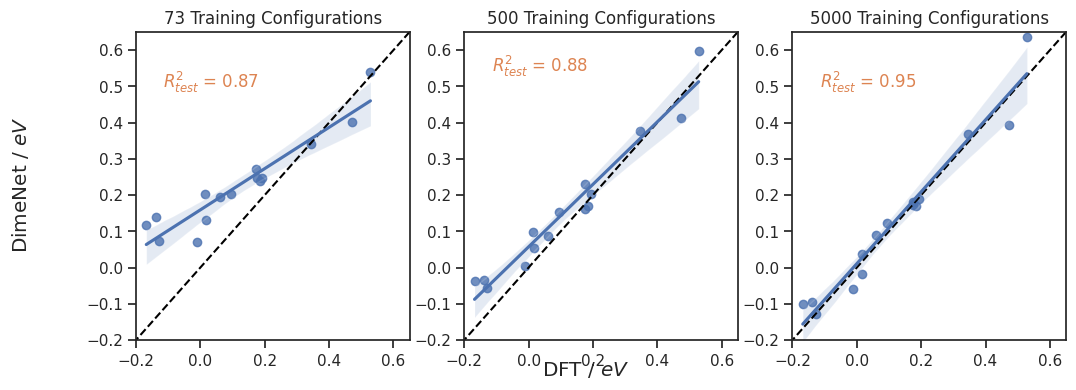}
    \end{subfigure}
    \begin{subfigure}{0.5\textwidth}
        \includegraphics[width=\linewidth]{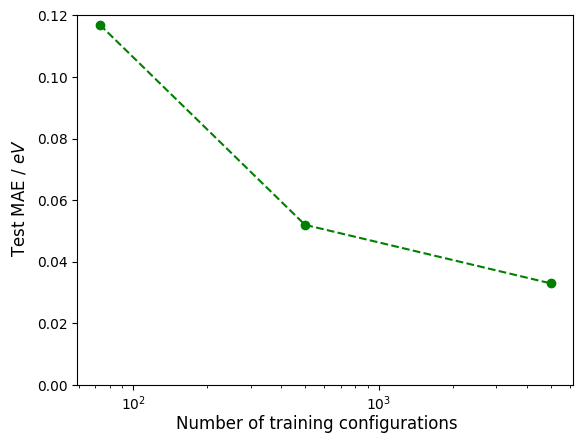}
    \end{subfigure}
    \caption{Test performance of DimeNet training with more data provided by cluster expansion model}
    \label{fig:cedata-gnn}
\end{figure}

Since cluster expansion model is superior in prediction accuracy and uncertainty, we use this cluster expansion model to generate more target energies with different surface configurations at low concentrations and support the training of DimeNet. Using the energies predicted by cluster expansion as surrogates,  it bypass the need of computational cost for thousands of supercell of DFT calculationsto test the potential of DimeNet applied at systems with interacting defects. We trained DimeNet with 500 and 5,000 CE generated energies respectively and the results are in Figure \ref{fig:cedata-gnn} and Table \ref{table2}. As there are more training data points, DimeNet gains more expressive representation. When training with 5,000 configuration, DimeNet is able to predict as good as MBTR designed by domain expert. This finding motivate us to do further studies on how DimeNet captures relationship between defects and interactions by utilizing more low concentration defect configurations and whether the representation generalize to high concentration systems.  

\begin{table}[h]
  \caption{DimeNet train on more configurations}
  \label{table2}
  \centering
  \begin{tabular}{llll}
    \toprule
    Number of Training Configurations  & 73 & 500 & 5000\\ 
    \midrule
    Test MAE ($eV$)   & 0.117 & 0.052    & 0.033\\
    \bottomrule
  \end{tabular}
\end{table}
\section{Conclusion}
In this work, we explored three approaches to predict the free energy change of surface systems with interacting defects with a small data set. While materials descriptor with classical machine learning algorithm performs well with limited data, cluster expansion is more accurate and less biased in prediction. Within the scope of interacting defects scattered on surface, cluster expansion model can serve as a means to provide training data at low cost to survey the potential of applying deep neural networks in the task of property prediction for surface with interacting defects. On the other hand, it is worth considering the necessity of using machine learning with descriptors or deep learning to study the defects interactions at surface only when cluster expansion model performs well with purchasable cost to obtain training data.

\section{Disclaimer}
This project was conducted independently during my early stage of graduate studies and may not have received formal supervision. If you come across this report, please feel free to contact me with any issues or suggestions you may have.

\medskip

\newpage
\bibliography{reference}
\end{document}